\documentclass[12pt,preprint]{aastex}

\def\Rsch           {\ifmmode {R_{\rm Sch}} \else  {$R_{\rm Sch}$} \fi}
\def\MoverR         {\ifmmode {M/R} \else  {$M/R$} \fi}
\def\rhoh           {\ifmmode {\rho_h} \else  {$rho_h$} \fi}
\def\Msunpercpc     {M$_\odot$~pc$^{-3}$}

\newcommand{\masyr}	{mas yr$^{-1}$}
\newcommand{\masyrsq}	{mas yr$^{-2}$}
\newcommand{\sgra}	{Sgr~A*}
\newcommand{\sgrab}	{Sgr~A*~}
\newcommand{\uas}	{\ifmmode {\mu{\rm as}}\else {$\mu$as} \fi}
\newcommand{\uasyr}	{\ifmmode {\mu{\rm as~yr}^{-1}}\else {$\mu$as~yr$^{-1}$} \fi}
\newcommand{\muell}	{\ifmmode {\mu_l}\else {$\mu_l$} \fi}
\newcommand{\mubee}	{\ifmmode {\mu_b}\else {$\mu_b$} \fi}
\newcommand{\tnot}	{\ifmmode {\Theta_0}\else {$\Theta_0$} \fi}
\newcommand{\rnot}	{\ifmmode {R_0} \else {$R_0$} \fi}
\newcommand{\onot}	{\ifmmode {\Omega_0}\else {$\Omega_0$} \fi}
\newcommand{\vsun}	{\ifmmode {V_\odot} \else {$V_\odot$} \fi}
\newcommand{\wsun}	{\ifmmode {W_\odot} \else {$W_\odot$} \fi}
\newcommand{\peryr}	{y$^{-1}$}

\newcommand{\porm}	{\ifmmode {~\pm~} \else {$~\pm~$} \fi}
\newcommand{\kms}	{\ifmmode {{\rm km~s}^{-1}} \else {km~s$^{-1}$} \fi}
\newcommand{\kmskpc}	{\ifmmode {{\rm km~s}^{-1} {\rm kpc}^{-1}} \else {km~s$^{-1}$ kpc$^{-1}$} \fi}
\newcommand{\kmsyrsq}	{\ifmmode {{\rm km~s}^{-1} {\rm yr}^{-1}} 
                         \else {km~s$^{-1}$ yr$^{-1}$} \fi}
\newcommand{\msun}	{\ifmmode {{\rm M}_\odot} \else  {${\rm M}_\odot$} \fi}
\newcommand{\Msun}	{\ifmmode {{\rm M}_\odot} \else  {${\rm M}_\odot$} \fi}
\newcommand{\lsun}	{\ifmmode {{\rm L}_\odot} \else  {${\rm L}_\odot$} \fi}
\newcommand{\vmax}	{\ifmmode {V_{\rm max}} \else {$V_{\rm max}$} \fi}
\newcommand{\dNdr}	{\ifmmode {dN \over dr} \else {$dN \over dr$} \fi}
\newcommand{\ie}	{i.e.,~}
\newcommand{\eg}	{eg,~}
\newcommand{\etal}	{et al.~}

\newcommand{\Mr}        {\ifmmode {M_r} \else {$M_r$} \fi}
\newcommand{\Mtotal}    {\ifmmode {M_{{\rm total}}} \else {$M_{{\rm total}}$} \fi}
\newcommand{\rmin}      {\ifmmode {r_{{\rm min}}} \else {$r_{{\rm min}}$} \fi}
\newcommand{\rmax}      {\ifmmode {r_{{\rm max}}} \else {$r_{{\rm max}}$} \fi}
\newcommand{\rhonot}    {\ifmmode {\rho_0} \else {$\rho_0$} \fi}
\newcommand{\dm}        {\ifmmode {\delta m} \else {$\delta m$} \fi}
\newcommand{\twopi}     {\ifmmode {2\pi}   \else {$2\pi$}   \fi}

\def\p              {\phantom{>}}
\def\q              {\phantom{0}}

\newbox\grsign \setbox\grsign=\hbox{$>$} \newdimen\grdimen \grdimen=\ht\grsign
\newbox\laxbox \newbox\gaxbox
\setbox\gaxbox=\hbox{\raise.5ex\hbox{$>$}\llap
     {\lower.5ex\hbox{$\sim$}}}\ht1=\grdimen\dp1=0pt
\setbox\laxbox=\hbox{\raise.5ex\hbox{$<$}\llap
     {\lower.5ex\hbox{$\sim$}}}\ht2=\grdimen\dp2=0pt
\newcommand{\gax}{\mathrel{\copy\gaxbox}}
\newcommand{\lax}{\mathrel{\copy\laxbox}}

\shorttitle{The Case for a Supermassive Black Hole}
\shortauthors{Reid \& Brunthaler}

\begin{document}

\title{The Proper Motion of Sagittarius A*: III. The Case for a Supermassive Black Hole}

\author{M.~J.~Reid}
\affil{Center for Astrophysics $|$ Harvard \& Smithsonian, Cambridge, MA 02138}
\email{reid@cfa.harvard.edu}

\author{A.~Brunthaler}
\affil{Max-Planck-Institut f\"ur Radioastronomie, Auf dem H\"ugel 69, 
 D-53121 Bonn, Germany}
\email{brunthal@mpifr-bonn.mpg.de}

\slugcomment{January 13, 2020}

\begin{abstract}
	We report measurements with the Very Long Baseline Array of the proper 
motion of \sgra\ relative to two extragalactic radio sources spanning 18 years.
The apparent motion of \sgra\ is 
$-6.411\pm0.008$~mas~\peryr\ along the Galactic plane and 
$-0.219\pm0.007$~mas~\peryr\ toward the North Galactic Pole.
This apparent motion can almost entirely be attributed to the effects 
of the Sun's orbit about the Galactic center.  Removing these effects 
yields residuals of $-0.58\pm2.23$ \kms\ in the direction of Galactic
rotation and $-0.85\pm0.75$ \kms\ toward the North Galactic Pole.
A maximum-likelihood analysis of the motion, both in the Galactic plane 
and perpendicular to it, expected for a massive object within the Galactic 
center stellar cluster indicates that the radiative source, \sgra, contains 
more than about 25\% of the gravitational mass of $4\times10^6~\msun$ deduced 
from stellar orbits.  
The intrinsic size of \sgra\ is comparable to its Schwarzschild radius, and the 
implied mass density of $\gax4\times10^{23}$~\msun pc$^{-3}$ is very close to
that expected for a black hole, providing overwhelming evidence that it is 
indeed a super-massive black hole.
Finally, the existence of ``intermediate mass'' black holes more massive 
than $\approx3\times10^4$~\msun\ between approximately 0.003 and 0.1~pc from 
\sgrab are excluded.
\end{abstract}


\section{Introduction}

At the Galactic center, stars orbit an unseen mass of $4\times10^6$ \Msun\
\citep[\eg][]{Boehle:16,Gillessen:17}, and the compact radio source \sgrab projects 
to within $\approx1$ mas ($\approx8$ AU) of the gravitational focal position 
\citep{MREG97,Reid:07} of these stars.  If we are to conclude that the Galactic
center harbors a supermassive black hole, a critical question is how much of the 
unseen mass can be directly tied to \sgra.  Since the luminosity of \sgrab is only 
comparable to a stellar source, other information is needed to establish if it is a 
supermassive black hole.  To that end, we have been measuring the position of \sgrab 
with the National Radio Astronomy Observatory's\footnote{The National Radio Astronomy 
Observatory is operated by Associated Universities Inc., under a cooperative agreement 
with the National Science Foundation.} Very Long Baseline Array (VLBA) since 1995, since
a very massive object at the dynamical center of the Galaxy should be nearly motionless.  

The apparent motion of \sgra, relative to extragalactic radio sources, 
contains the reflex of the Sun's velocity in its orbit about the Galactic center, 
plus any intrinsic motion of \sgrab itself.  
In \citet{R99} and \citet{Reid:04}, hereafter Papers I and II, we published results 
from the first 8 years of observation. We showed that the component of the apparent 
motion of \sgra\ {\it perpendicular} to the Galactic plane could be explained by 
the motion of the Sun toward the North Galactic Pole, limiting the intrinsic motion 
of \sgra\ to $\lax1$ \kms\ in one dimension.  Since a massive object embedded in a 
dense stellar cluster suffers gravitational Brownian motion and reaches thermal equilibrium 
with the perturbing stars \citep{CHL02}, the observed lack of motion for \sgrab 
provided a lower limit of $\sim0.4\times10^6$ \Msun for \sgrab (Reid \& Brunthaler 2004).   
This clearly associated a very large mass with the radiative source \sgra, and  
greatly strengthened the already strong case for \sgrab being a super-massive black hole 
(SMBH).
  
In this paper, we report new observations which now span 18 years, reducing
proper motion uncertainties by a factor of three to less than $\pm10~\uasyr,$ 
both in and out of the Galactic plane.  Coupling these results with independent 
measurements of the angular motion of the Sun in its orbit about the Galactic center, 
we are now able to use two dimensions of velocity information to provide a stronger 
and more robust lower limit for the mass of \sgra, significantly increasing confidence 
that it is indeed a black hole.

\section{Observations and Results}

Our observations using the National Radio Astronomy Observatory's VLBA started in 
1995 and have now continued to 2013.  Paper I reported early results 
for observations from 1995 through 1997, and observations through 2003 were reported
in Paper II.  Here we present new observations conducted in in 2007 and 2013 under
VLBA programs BR124 and BR173.  As VLBI technology progressed we increased the recorded
data rates.  For BR124 we observed with eight 8-MHz bands with Nyquist sampling and 
2-bits per sample for a total sampling rate of 256 Mb~s$^{-1}$.  The observations
spanned 8 hours and we placed three geodetic-like blocks at the beginning, middle
and end of the tracks in order to measure and remove tropoposheric and clock delays.  
We switched between sources every 15 seconds, using \sgrab as the phase-reference
for the background sourcews.  For BR173 we observed with 16 32-MHz bands with Nyquist 
sampling and 2 bits-per-sample for a total sampling rate of 2 Gb~s$^{-1}$. These 
observations spanned 6 hours and we placed four geodetic-like blocks evenly spaced throughout
the observations, and we switched between sources every 17 seconds.  Details of the
calibration procedures can be found in Papers I and II.  

After calibraiton, we imaged all sources and measured their positions by fitting
elliptical Gaussian brightness distributions. 
Table \ref{table:positions} lists all of our position measurements of \sgrab relative 
to two compact extragalactic radio sources, J1745--2820 and J1748--2907, in J2000 
Equatorial and also Galactic coordinates.  The measurements were made at the highest 
astrometrically useful frequency of the VLBA of 43 GHz in order to minimize the effects 
of strong interstellar scattering toward the Galactic center.\footnote{
While some antennas of the VLBA have 86 GHz receivers, the system sensitivity is 
approximately a factor of five poorer than at 43 GHz.  In addition, interferometer
coherence times are a factor of two shorter at 86 GHz compared to 43 GHz.  These factors
strongly favor 43 GHz observations requiring phase-referencing needed for astrometry.}
Position uncertainties include estimates of systematic effects, dominated by small residual 
errors in modeling atmospheric delays. 

\subsection{Proper Motion of \sgra}

\begin{figure}
\epsscale{1.0}
\plottwo{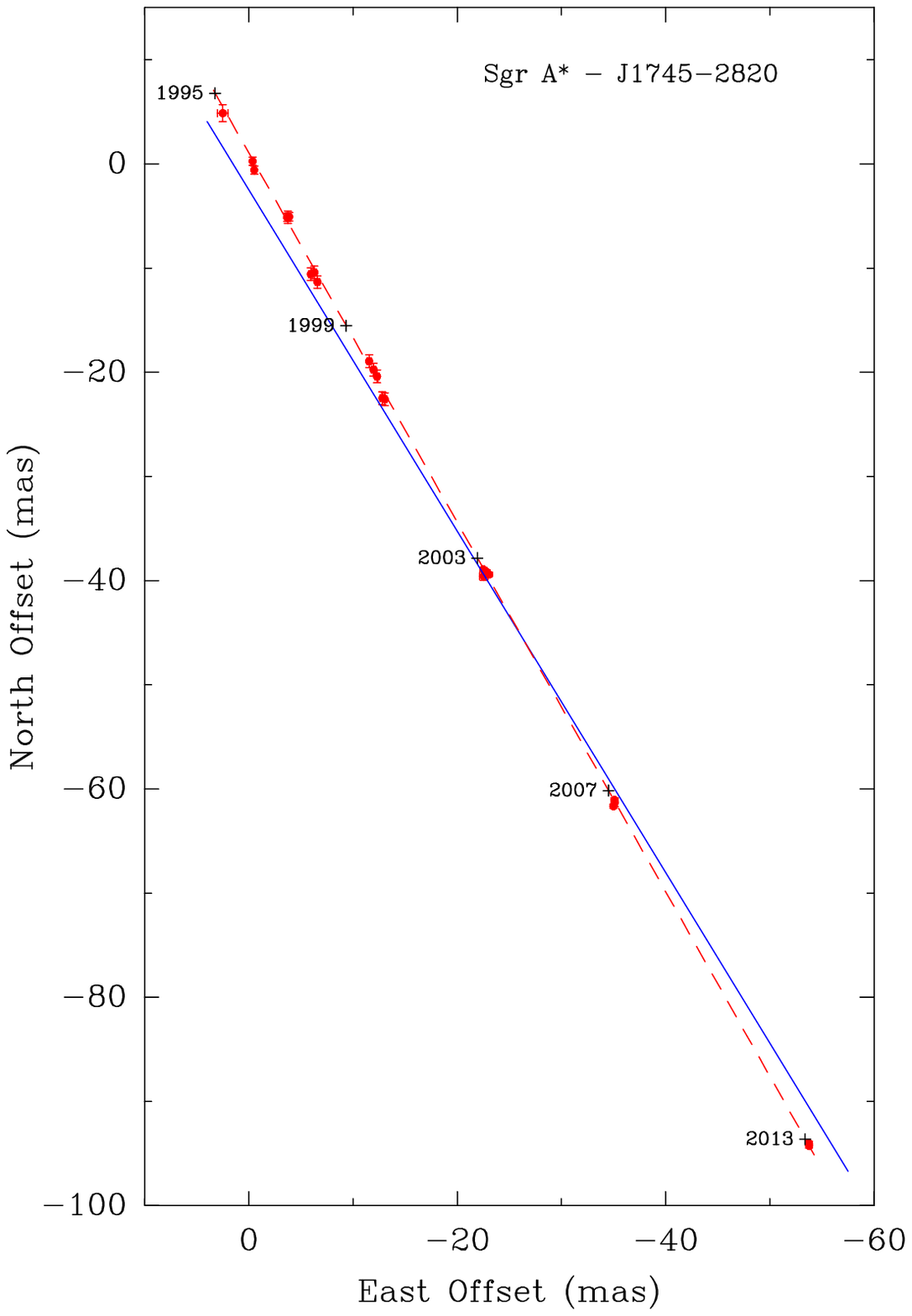}{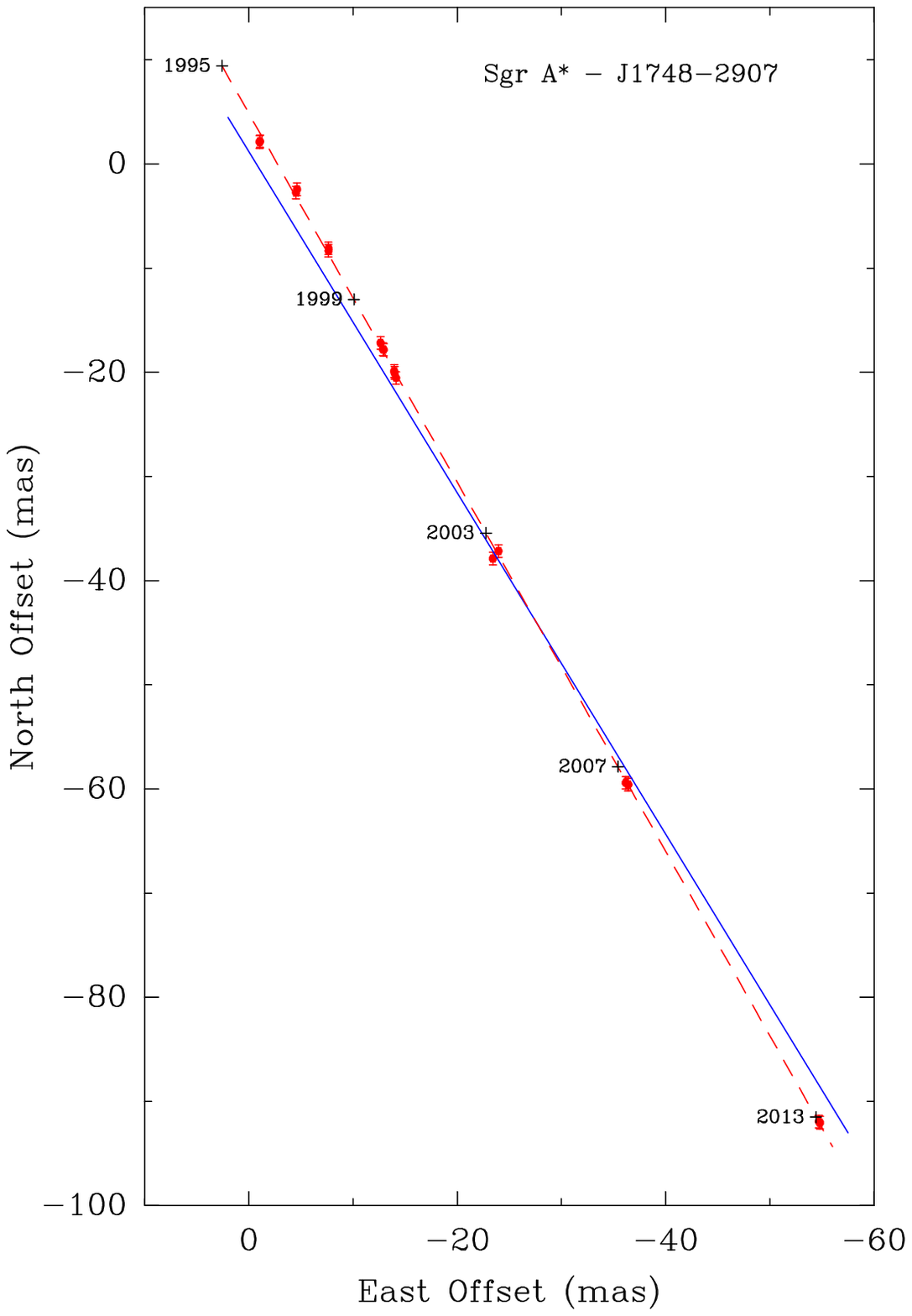}
\caption{\small
Position residuals of \sgrab\ relative to J1745--2820 (left panel) and J1748-2907
(right panel) on the plane of the sky.  
Measurements are indicated with ellipses, whose sizes are the
scatter-broadened image of \sgrab at 43 GHz, and $1\sigma$ error bars which are
dominated by systematic uncertainties.
The dashed lines are least-squares fitted proper motions;
the solid line gives the orientation of the Galactic plane when looking toward
the Galactic center.
 \label{fig:skyplots}}
\end{figure}

The positions on the sky of \sgra, relative to the two background sources, 
are plotted in Fig.~\ref{fig:skyplots}.  The observations now span 18 years and the
linear trends reported in Papers I and II continue.   Variance-weighted least-squares 
fits to the position versus time of \sgrab relative to J1745--2820 and J1748--2907
are given in Table \ref{table:motions} and plotted with dashed lines in Fig.~\ref{fig:skyplots}.
The results for the two background sources are consistent and differencing the motions
with respect to the background sources shows no significant motion.
Assuming that the background sources are sufficiently distant that they have negligible 
intrinsic angular motion, we averge the two results and estimate \sgra's apparent motion 
to be $-3.156\pm0.006$ and $-5.585\pm0.010$~\masyr\  in the easterly and northerly 
directions, respectively.  

\begin{figure}
\epsscale{1.0}
\plottwo{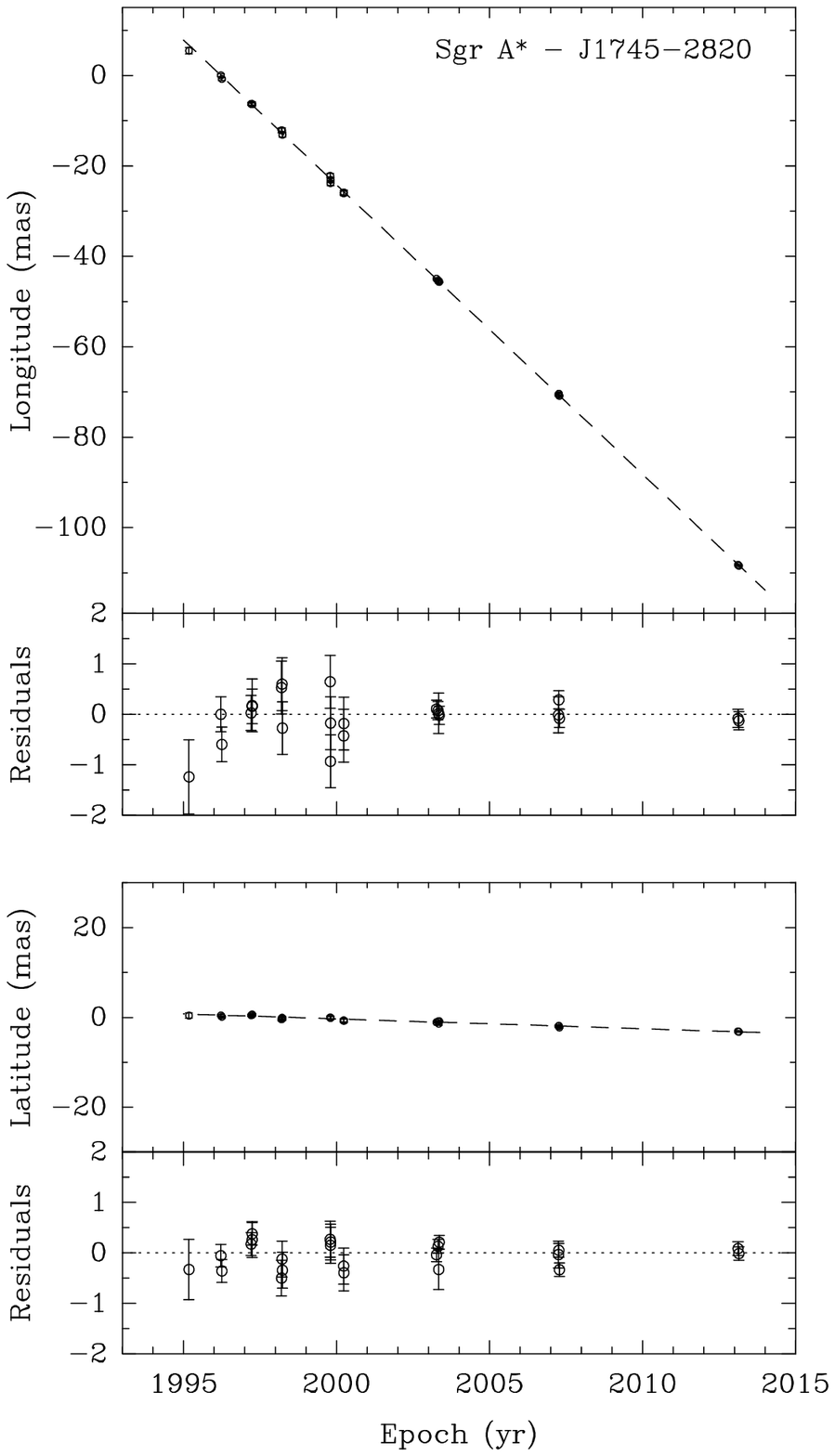}{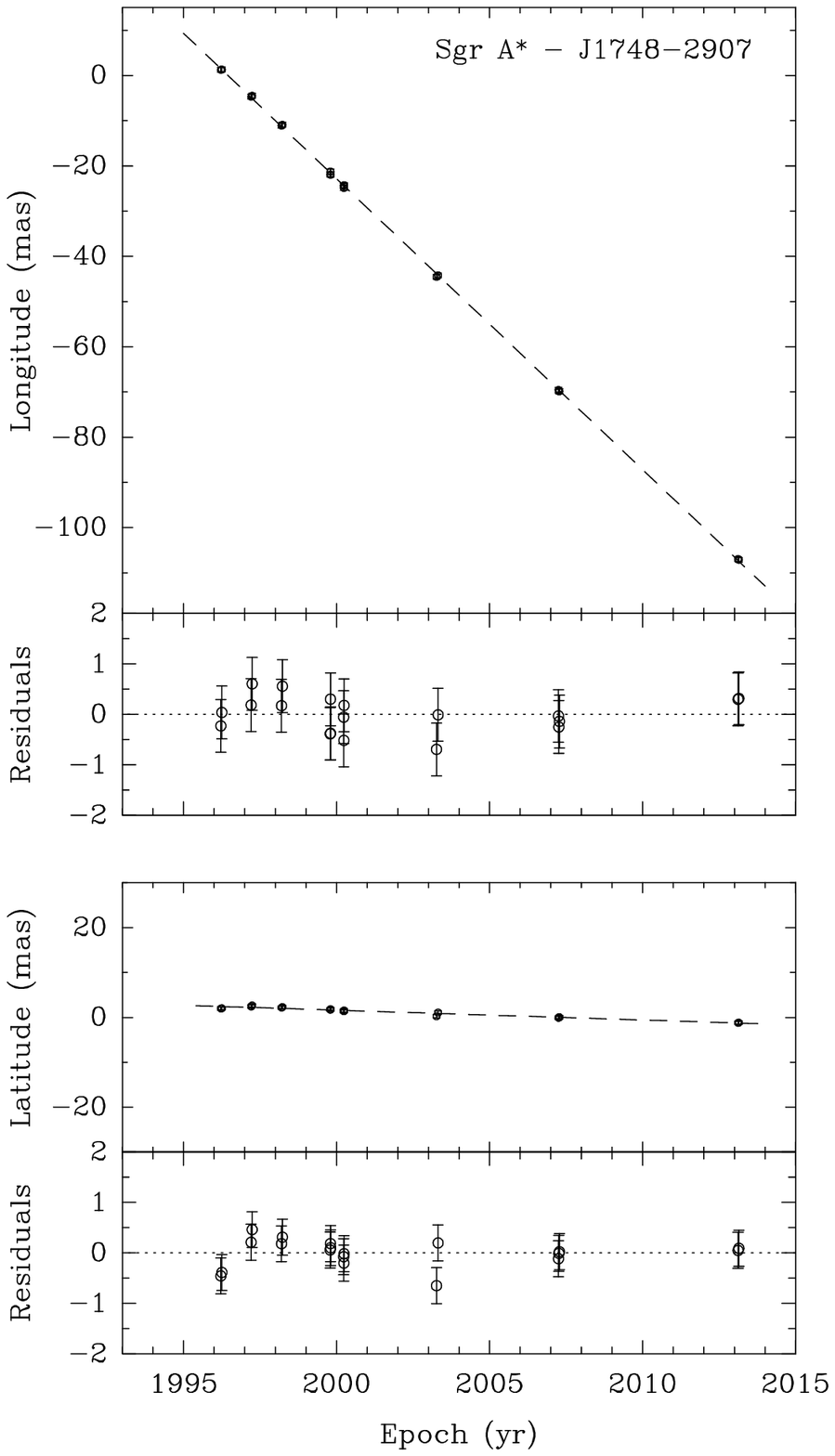}
\caption{\small
Galactic longitude and latitude vs time for \sgrab relative to J1745--2820 (left panels)
and J1748--2907 (right panels).  Dashed lines are variance-weighted least-squares fits to 
the data, and the residuals to those fits are shown below each plot. 
 \label{fig:galactic}}
\end{figure}

We re-fitted for motions in Galactic coordinates (see Table \ref{table:motions}), yielding 
apparent motion components for \sgrab in Galactic longitude of $-6.411\pm0.008$ \masyr\ 
and in latitude of $-0.219\pm0.007$ \masyr.  The data and fits are displayed in 
Fig. \ref{fig:galactic}.  Adopting a distance to the Galactic center of $\rnot=8.15$ kpc
\citep{Gravity:19,Do:19,Reid:19}, \sgrab {\it appears} to be moving predominantly along the 
Galactic plane with a tangential (toward increasing longitude) speed of $-247.69\pm0.33$ \kms\ 
and toward the North Galactic Pole with a speed of $-8.45\pm0.26$ \kms.

\subsection {Acceleration of \sgra}

We also investigated the possibility that \sgrab is being accelerated by, for 
example, an intermediate mass black hole (IMBH).  When we added an acceleration 
parameter in the motion fits, we obtain easterly and northerly accelerations estimates of
$-0.0026\pm0.0030$ \masyrsq\ and $-0.0050\pm0.0038$ \masyrsq\ relative to J1745--2820 and 
$0.0058\pm0.0029$ \masyrsq\ and $0.0129\pm0.0074$ \masyrsq\  relative to J1748--2907.
A variance-weighted average of the two acceleration results gives easterly and
northerly motions of $0.0017\pm0.0021$ \masyrsq\ and $-0.0013\pm0.0034$ \masyrsq.
These acceleration estimates are an order of magnitude improvement over our results
in Paper II.   They are consistent with no measurable acceleration, with a $2\sigma$ 
upper limit of 0.0080 \masyrsq\ (0.31 \kmsyrsq) for the magnitude of the two-dimensional 
acceleration vector.  Acceleration limits are potentially interesting as they require 
no correction for Solar orbital acceleration, which is $\sim10^{-7}$ \masyrsq\ \citep{GR98}. 
 
\begin{figure}
\epsscale{0.65}
\plotone{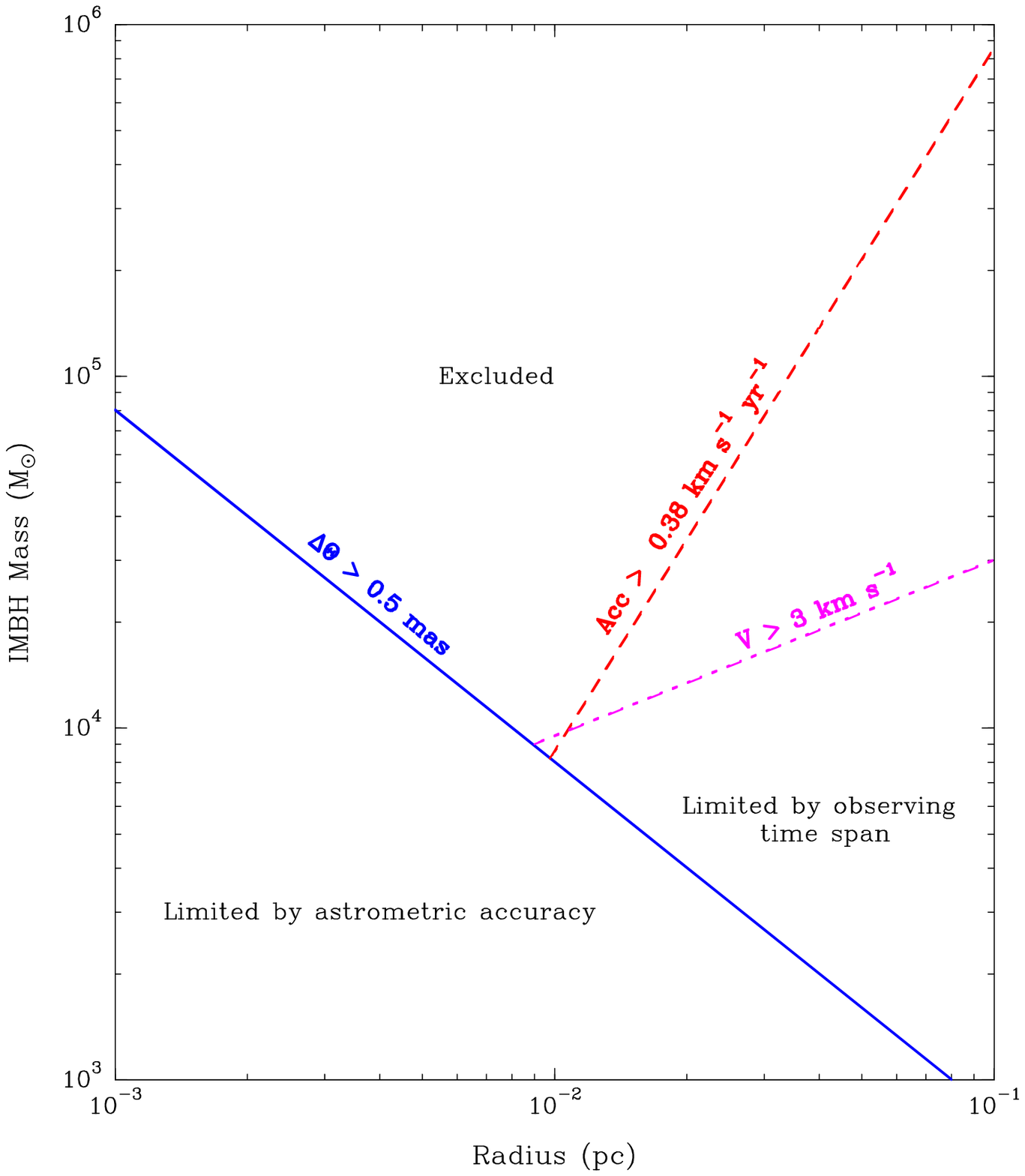}
\caption{\small
Limits on the mass of an intermediate mass black hole (IMBH) orbiting a $4\times10^6$ \msun
\sgrab as a function of orbital radius.  The blue line divides the parameter space for 
angular excursion of \sgrab of 0.5 mas, below which it could appear as jitter within our 
current astrometric accuracy.  The dashed red line corresponds to our upper limit on the (2-D) 
magnitude of the acceleration of \sgra, and the dot-dashed magenta line corresponds to similar
limits on the velocity of \sgra.  The region excluded by these 
constraints is above the dashed lines, which could be improved by a longer time span of 
observations.
 \label{fig:IMBH}}
\end{figure}

Interestingly, the stellar cluster IRS 13E has marginal evidence for an IMBH of 
$3 \times 10^4$ \Msun\ \citep[\eg][]{Genzel:10}.  The cluster projects within
$\approx3$ asec of \sgra, corresponding to a linear offset of $\approx0.1$ pc.
Assuming its three-dimensional distance from \sgrab is comparable to this offset, 
such an IMBH would likely induce an acceleration of $0.4~\kmsyrsq$ and a linear motion 
of about 3 \kms\ for \sgra. Fig. \ref{fig:IMBH} presents the regions of IMBH mass and 
radius from \sgrab that are allowed for and excluded by the current observations. 
Our limits make an IMBH of $\approx3\times10^4$~\msun\ between  0.003 and 0.1~pc from 
\sgrab unlikely and strongly exclude a more massive object of $\gax10^5~\msun,$ as does 
dynamical modeling of the orbit of star S02 (S2) by \citet{Naoz:19}.
Note that a modest increase in the time span of our astrometric observations could better 
test the IMBH's existence, since for uniform sampling acceleration accuracy improves as 
the 5/2 power and motion accuracy improves as the 3/2 power of time spanned.

\section {Limits on the Mass of \sgra}\label{sect:limits}

Some of the so-called ``S'' stars have been seen projected within $\sim0.001$ pc of 
\sgrab and move at thousands of \kms\ \citep{S02,Ghez05} as they orbit a dark mass 
concentration.  
In contrast, \sgra, which is located within 0.00004 pc of the gravitation focus
of the orbiting stars \citep{MREG97,R03,Reid:07}, has an intrinsic motion (relative to 
distant quasars) of less than a few \kms, strongly suggesting that it is very massive.  
For example, were \sgrab only a 10 \Msun black hole in an X-ray binary,
which would be consistent with its meager luminosity, it too should be moving at a great speed.
However, a massive object in the presence of large numbers of stars experiences 
gravitational Brownian motion, which is expected to result in equipartition of kinetic 
energy between the massive object and individual stars \citep{CHL02,Dor03,Merritt:07}.  
In Paper II, we performed detailed simulations of the motion of a massive object orbited 
by $\sim10^6 - 10^7$ stars within its sphere of influence.  These simulations confirmed 
that equipartition of kinetic energy is indeed achieved, and that a $4\times10^6$ \Msun\ 
object would be expected to have a (one-dimensional) motion of between 0.18 and 0.30 \kms, 
depending on the nature of the stellar mass function.  In addition, a cluster of dark 
stellar remnants summing to $0.4\times10^6$ \Msun\ could contribute an additional 0.2 \kms\ 
to the motion of the massive object.  In this context, an upper limit to 
\sgra's intrinsic motion can provide a lower limit to its mass. 

Compared to Paper II, our latest results for the apparent motion of \sgrab have decreased 
the motion uncertainties by a factor of three.  Also, in Paper II we used only the 
component of motion of \sgrab perpendicular to the Galactic plane, corrected for the Sun's 
motion, to provide a lower limit for the mass of \sgra.  The motion of \sgrab in the 
Galactic plane was not used since, at the time, the correction for the orbital motion of 
the Sun was quite uncertain ($\pm20$ \kms).  The Sun's motion in its Galactic orbit is now 
known to much higher accuracy, and we can now use the Galactic longitude motion to 
complement the latitude motion in order to more tightly and robustly constrain the mass of 
\sgra. 

By modeling parallaxes and proper motions of 
about 150 massive young stars with maser emission, \citet{Reid:19} estimate the 
angular speed of the Sun in its Galactic orbit with sub-percent accuracy: 
$(\tnot+\vsun)/\rnot = 30.32\pm0.27$ \kmskpc, where $\tnot$ is the circular orbital
speed in the Galaxy at the Sun and $\vsun$ accounts for the Sun's deviation
from a perfect circular orbit.   In Table \ref{table:intrinsicmotion},
we detail the values used to remove the effects of the Sun's motion, in order
to estimate the intrinsic motion of \sgra, both in and out of the Galactic plane.  
We find that \sgrab is nearly motionless with longitude and latitude speeds of
$-0.58\pm2.23$ and $-0.85\pm0.75$ \kms, respectively.  Since these speeds came from
differencing independently determined angular motions, the adopted value of \rnot\ of 8.15 kpc 
used to convert the differences to linear motions appears only as a final scale factor, and
since \rnot\ is now known to better than 2\% accuracy \citep{Gravity:19,Do:19,Reid:19}, its
uncertainty is not important for our application.

In order to estimate how massive is \sgrab, we simulate the effects of the central
star cluster on a massive central object.
Using the same approach as in Paper II, we generate $10^3$ random configurations of
stars orbiting an object with a given trial mass, follow the system for the time spanned 
by our observations (now 18 years), and infer the motion of the trial object
from the change in center of mass of the orbiting stars.  We compare the simulated components
of motion in each of two dimensions with trials drawn from Gaussian distributions, which
are consistent with our observed intrinsic motion of \sgrab in Galactic longitude and
latitude.  We keep track of the fraction of trials that give at least one component of the 
massive object's velocity which is inconsistent with our observed limits.  
We then repeat the simulation with different trial masses in order to trace the 
distribution as a function of mass.  In contrast to Paper II, which only used 
the latitude motion of \sgrab to compare to the simulations, 
we now use two components (latitude and longitude) to better constrain \sgra's mass.  
This significantly improves both the lower limit on the mass of \sgrab and the 
robustness of the mass estimate.

In Paper II we evaluated three stellar initial mass functions (IMFs): a standard 
IMF, a top-heavy IMF with a high-mass index by flatter by 0.5, and one flatter by 1.0.
Given strong evidence for a top-heavy IMF in the Galactic center region, but
an uncertain flattening at high masses \citep[\eg][]{F99,SGB02,Genzel:10}, 
we conservatively adopt the moderately flattened IMFs considered in Paper II (with
an index flatter by 0.5).   We assume the broken power-law radial distribution of stars 
given by Eq. 4 of \citet{Genzel:10}.  Specifically, our fiducial model has a volume 
density of stars given by 
$\rho_*(R) = 1.35\times10^6~(R/0.25 {\rm pc})^{-\gamma}$ $\msun {\rm kpc}^{-3}$, 
with $\gamma=1.3$ and $1.8$ inside and outside of $R=0.25$ pc, respectively.  We then generate
random orbital parameters with semi-major axes between $R=100$ AU (approximately
the smallest radius observed for stars) and $2.9$ pc (corresponding to the radius of the 
sphere of influence of a $4\times10^6$ \msun central mass in the Galactic center).
 
\begin{figure}[ht]
\epsscale{0.85}
\plotone{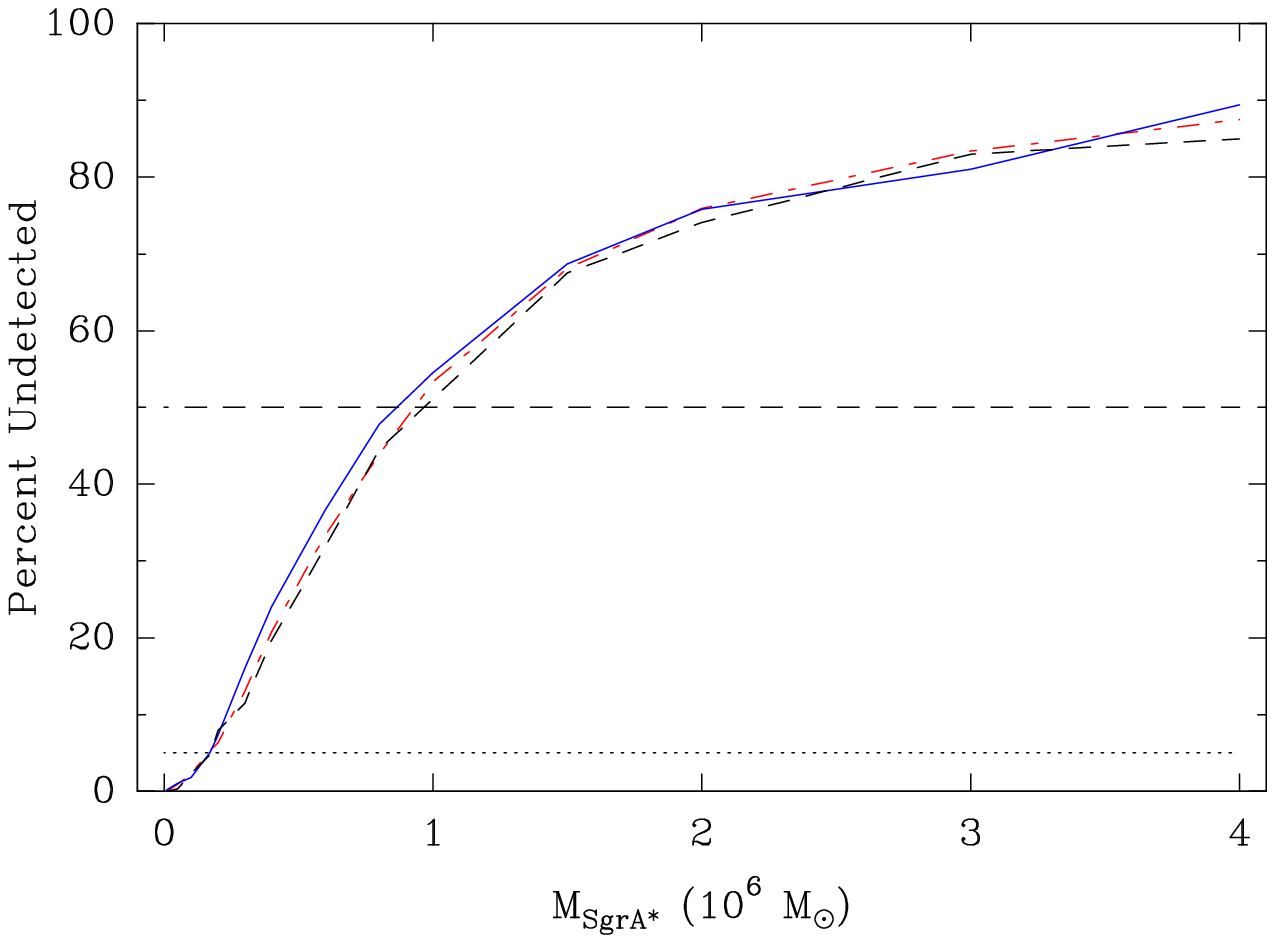}
\caption{\small
Results of simulations of the motion of \sgra, owing to random perturbations
from the central stellar cluster, compared to observed limits.  
Plotted are the percentages of simulations that fall below the observed limits 
as a function of the assumed mass of \sgra.  All simulations
assume a stellar mass function index flatter by 0.5 compared to a standard Salpeter IMF.
The blue solid line is our fiducial result, which uses the updated stellar distributions 
from \citet{Genzel:10} with stellar density $\rho_*(R) = 1.35 \times 10^6~(R/R_b)^{-\gamma},$
where $R_b=0.25$ pc is a break radius and $\gamma$ is 1.3 for $R<R_b$ and 1.8 for larger radii.
The black dashed line adds $\approx4,000$ stellar-mass black holes with a Bahcall-Wolff
cusp-like radial distribution with index $-7/4$ in the inner 0.2 pc.
For comparison, the red dot-dashed line uses the stellar distribution from Paper II.
Dotted and dashed lines indicate 5\% and 50\% of simulations that would not have been detected, 
corresponding to \sgrab masses of about $0.2$ and $1.0\times10^6~\Msun,$ respectively.   
 \label{fig:sgramass}
       }
\end{figure}

In Fig. \ref{fig:sgramass} we plot the results of these simulations with a solid blue line.
We find that half of the trials would be detected were \sgrab less massive than 
$0.8\times10^6$ \Msun, a factor of two stronger limit than in Paper II.  Importantly,
our result for 95\% confidence (\ie 5\% undetected) is now $0.2\times10^6$ \Msun,
a factor of 200 improvement over the result in Paper II.  This improvement comes mostly 
from the use of two dimensions of motion, compared to the one dimension available in Paper II.
For comparison, we repeated the simulations using the stellar cluster model employed
in Paper II and plot the results with a red dash-dotted line in Fig. \ref{fig:sgramass}.
The differences between the stellar cluster models results only in small differences
in the final results.   

Since one expects a significant population of stellar remnants to accumulate 
in the Galactic center, we added a population of stellar-mass black holes to the
fiducial model of the central cluster and re-ran the simulations.  
Following the models of \citet{Freitag:06}, we added a total mass of $7\times10^4$ \msun\  
in black holes between radii of 100 AU and 0.2 pc.  This corresponds
to less that 2\% of the mass within 0.2 pc of the center.  We assumed a flat 
distribution in mass between 6 and 30 $\msun,$ which results in about 4,000 black holes, 
and a cusp-like radial distribution with a Bahcall-Wolff power-law index of $-7/4$. 
The results of these simulations are shown with
the black dashed line in Fig. \ref{fig:sgramass}.  As expected, the motions of 
\sgrab increase, but only modestly, and 50\% of the trials would be detected were 
\sgrab to hold $1.0\times10^6$ \msun.

Note that all of our simulations use smooth distributions of stars, without any clumping.  
Since clumping would increase the simulated motion of \sgra, our mass limits are very 
conservative.  Thus, we adopt a round-number mass of $1.0\times 10^6$ \Msun\ as a 
maximum-likelihood lower limit for the mass of the radiative source, \sgra, when we explore 
its significance in the next section.

\section {Is \sgrab a Super-Massive Black Hole}\label{sect:SMBH}

Given that the radiative source \sgrab likely has a mass greater than 
$10^6~\Msun,$ how does this help answer the question ``is it a black hole''?
If one can show that sufficient mass is contained within a small enough volume, 
Einstein's theory of General Relativity requires a black hole.
This leads to a maximum mass density that matter can achieve
before a black hole forms. Thus, the case for the existence of supermassive 
black holes centers on observations of objects that approach a critical mass 
density. The size of a black hole of a given mass can be defined by its 
Schwarzschild radius, $R_{Sch} = 2 G M/c^2~~,$ 
where $G$ is the gravitational constant, $M$ the mass of the black hole, 
and $c$ is the speed of light.
However, any matter that comes within $3R_{Sch}$ of a (non-rotating) hole cannot 
achieve a stable orbit and falls directly into the black hole.
Thus, a critical mass density, $\rho_{crit}$, to require a black hole is
the mass divided by the volume enclosed by $3R_{Sch}$:
$$\rho_{crit} = {M \over (4\pi/3)(3R_{Sch})^3}~~.$$
Substituting the relation for $R_{Sch}$, we find
$$\rho_{crit} = {c^6\over 288\pi G^3 M^2}~~.$$
Note that this density depends inversely as the square of the mass.\footnote{
Interestingly, the critical density reaches that of water for a 30-million solar mass  
black hole, $10^{15}$ times lower than for a 1 solar mass black hole.  Thus, for example, 
an ``ocean'' contained within the orbit of Mars would be a black hole.}

An alternative ``linear'' density, $\phi_{crit}$, defined as $M/3R_{Sch}$
provides a simple mass-independent parameter that can be used to establish a black hole:
$$\phi_{crit} = {2 G \over 3c^2}~~.$$

Table \ref{table:densities} lists critical linear and volume density limits as established 
from various observations and updated from \citet{Reid:09}.    
The first table entry is for globular clusters, which can have 
upwards of $\sim10^6$ stars within a radius of $\sim1$ pc.  
This provides a reference point for relatively high stellar densities that are commonly 
achieved in galaxies.  The last table entry gives the critical linear and volume densities 
for a theoretical SMBH of $4\times10^6$ \Msun\ assuming a radius of $3R_{Sch}$. 
If one can show that these critical densities are achieved, the case for a supermassive
black hole is established with near certainty.

Now we consider systems only with enclosed mass estimates from well-defined Keplerian orbits. 
The second entry in Table \ref{table:densities} is based on observations of water masers in
the center of the galaxy NGC 4258.  Very Long Baseline Interferometric (VLBI) observations 
with angular resolution of 0.4 mas show that these masers originate from a slightly 
warped, thin disk of gas within an angular radius of about 3 mas ($2\times10^4$ AU 
at the distance of the galaxy).  The observed rotational speeds of 900 \kms\ are consistent 
with Keplerian orbits about a central mass of $4\times10^7$ \Msun.  The implied volume 
density is nearly five orders of magnitude above stellar densities in globular clusters, 
effectively ruling out a cluster of normal
stars\footnote{These observations, and those for dozens of other galaxies, certainly 
provide strong evidence that supermassive black holes are found at the centers of active 
galaxies.  Were one to place $\sim10^7$ stars inside a radius of $\sim0.1$ pc, the system 
would be dynamically unstable.   Less massive stars would be expelled while dynamical friction
would cause massive stars to sink to the center where they could possibly form a black hole.  
These observations rule out long-lived clusters of normal stars as providing the central 
gravitational mass \citep{M98}.  However, this conclusion is based on the 
assumption of an {\it isolated} stellar cluster. The possibility of a quasi-steady-state 
condition, wherein stars beyond the 0.1 pc radius are gravitationally perturbed and enter 
the central region, replenishing those expelled, has yet to be considered in detail.}.  
However, it is not sufficient to rule out clusters of compact stellar 
remnants (eg, white dwarfs, neutron stars, or stellar black holes).

The third entry in Table \ref{table:densities} is based on infrared observations
of stars that orbit an unseen mass in the center of the Milky Way.  These observations by 
the Max Planck Institue for Extraterrestrial Physics \citep[\eg][]{Gillessen:17} and 
the Univeristy of California, Los Angeles \citep[\eg][]{Boehle:16} groups require a 
central mass of $4\times10^6$ \Msun within a radius of 100 AU.  One star (S2; S02) has 
been seen to complete nearly two elliptical orbits and multiple stars have traced partial 
orbits.  All stars show a common gravitational focal position, which
coincides with the radio source Sgr A* to within $\approx0.001$ arcseconds
\citep{MREG97, Reid:07}, and require the same central mass.  The inferred mass density
of $>8\times10^{15}$ \Msunpercpc\ is high enough to rule out very long-lived
clusters of stars, as well as a speculative proposal of a central ``ball'' of heavy 
fermions \citep{Munyaneza:02}.   

The fourth entry in Table \ref{table:densities} refers to the VLBI observations of the 
proper motion of Sgr A* reported in this paper. As shown in Section \ref{sect:limits}, 
 Sgr A* {\it appears} to move along the plane of the Milky Way in a manner that can be 
completely accounted for by our orbit about the center of the Milky Way.  This 
provides an upper limit of $\sim1$ \kms\ for the intrinsic motion of Sgr A* itself.
Stars near Sgr A* have been observed to move at thousands of \kms, and the only way that 
Sgr A* can be motionless is for it to be extremely massive.  Both theory and direct 
simulations of the gravitational ``Brownian'' motion of a supermassive object at the 
center of the observed stellar cluster require \sgrab to be in near thermal equilibrium 
with the stars within its sphere of influence.  Our detailed simulations of the effects
of the central stellar cluster on the expected motion of \sgra, described in Section 
\ref{sect:limits}, constrain its mass to likely exceed $1\times10^6$ \Msun.   
A long history of VLBI observations have gradually improved measurements of the intrinsic 
size of Sgr A*; the most recent show Sgr A*'s emission has a radial extent of 
$\approx0.18$ AU \citep{D08}\footnote{The measured 
apparent size of Sgr A* is slightly smaller than that given by $3R_{Sch}$.
This is as expected for the radiation from material in a disk orbiting the black hole at 
$3R_{Sch}$.  Since the approaching material on one side is moving toward us at nearly 
the speed of light, this emission is boosted by relativistic aberration and Doppler shifts, 
causing this side to dominate the emissions.}, which is comparable to the Schwarzschild radius 
for a $4\times10^6$ \Msun\ black hole.  

Combining the lower limit for Sgr A*'s mass
(from its lack of motion) with the size of the source yields both linear and volume mass 
densities that are within a factor of about three of the General Relativity limit for 
a black hole.  This provides overwhelming evidence for a supermassive black hole
at the center of the Milky Way.

\section {Conclusions}

In summary, infrared observations of stars orbiting an unseen mass concentration provide 
extremely strong evidence for a supermassive black hole 
at the center of the Milky Way.  Radio observations associate that unseen mass
with the radiative source \sgra, and its lack of motion requires a huge mass to 
reside within a region of a few Schwarzschild radii.  If, following the infrared 
observations, there was any doubt that Sgr A* is a supermassive black 
hole, the radio observations should remove that doubt.

{\it Facilities:}  \facility{VLBA}

\acknowledgments

\clearpage

\begin{deluxetable}{ccrrrrr}
\tablewidth{0pt}
\tabletypesize{\small}
\tablecaption{Residual Position Offsets Relative to \sgra}
\tablehead{
\colhead{Source}         & \colhead{Date of}      &
\colhead{East Offset}& \colhead{North Offset}&
\colhead{$\ell^{II}$ Offset}    & \colhead{$b^{II}$ Offset} &
 \\
\colhead{}               & \colhead{Observation}   &
\colhead{(mas)}       & \colhead{(mas)    }  &
\colhead{(mas)}       & \colhead{(mas)    }  &
 }
\startdata
J1745--283  & 1995.178 &$  -2.50\porm 0.5$ &$  -4.87\porm 0.8$ &$ -5.46\porm0.73$ &$-0.40\porm0.60$ \\
            & 1996.221 &$   0.37\porm 0.1$ &$  -0.25\porm 0.4$ &$ -0.02\porm0.34$ &$-0.44\porm0.22$ \\
            & 1996.252 &$   0.52\porm 0.1$ &$   0.59\porm 0.4$ &$  0.77\porm0.34$ &$-0.14\porm0.22$ \\
            & 1997.211 &$   3.67\porm 0.1$ &$   5.13\porm 0.4$ &$  6.29\porm0.34$ &$-0.46\porm0.22$ \\
            & 1997.241 &$   3.87\porm 0.1$ &$   5.08\porm 0.4$ &$  6.35\porm0.34$ &$-0.66\porm0.22$ \\
            & 1997.241 &$   3.76\porm 0.2$ &$   5.13\porm 0.6$ &$  6.33\porm0.52$ &$-0.54\porm0.36$ \\
            & 1998.202 &$   5.95\porm 0.2$ &$  10.58\porm 0.6$ &$ 12.13\porm0.52$ &$ 0.43\porm0.36$ \\
            & 1998.219 &$   6.29\porm 0.2$ &$  10.42\porm 0.6$ &$ 12.17\porm0.52$ &$ 0.06\porm0.36$ \\
            & 1998.230 &$   6.59\porm 0.2$ &$  11.34\porm 0.6$ &$ 13.11\porm0.52$ &$ 0.28\porm0.36$ \\
            & 1999.791 &$  11.55\porm 0.2$ &$  18.95\porm 0.6$ &$ 22.19\porm0.52$ &$ 0.01\porm0.36$ \\
            & 1999.799 &$  12.29\porm 0.2$ &$  20.40\porm 0.6$ &$ 23.82\porm0.52$ &$ 0.13\porm0.36$ \\
            & 1999.805 &$  11.97\porm 0.2$ &$  19.75\porm 0.6$ &$ 23.10\porm0.52$ &$ 0.07\porm0.36$ \\
            & 2000.232 &$  13.04\porm 0.2$ &$  22.60\porm 0.6$ &$ 26.08\porm0.52$ &$ 0.64\porm0.36$ \\
            & 2000.238 &$  12.82\porm 0.2$ &$  22.49\porm 0.6$ &$ 25.88\porm0.52$ &$ 0.77\porm0.36$ \\
            & 2003.264 &$  22.51\porm 0.1$ &$  38.94\porm 0.2$ &$ 44.96\porm0.18$ &$ 1.08\porm0.14$ \\
            & 2003.318 &$  22.84\porm 0.1$ &$  39.18\porm 0.2$ &$ 45.34\porm0.18$ &$ 0.92\porm0.14$ \\
            & 2003.339 &$  22.54\porm 0.4$ &$  39.59\porm 0.4$ &$ 45.53\porm0.40$ &$ 1.38\porm0.40$ \\
            & 2003.353 &$  23.06\porm 0.1$ &$  39.41\porm 0.2$ &$ 45.66\porm0.18$ &$ 0.85\porm0.14$ \\
            & 2007.253 &$  35.13\porm 0.2$ &$  61.29\porm 0.4$ &$ 70.62\porm0.36$ &$ 1.95\porm0.27$ \\
            & 2007.264 &$  35.08\porm 0.1$ &$  61.06\porm 0.2$ &$ 70.39\porm0.18$ &$ 1.87\porm0.14$ \\
            & 2007.280 &$  35.00\porm 0.1$ &$  61.66\porm 0.2$ &$ 70.86\porm0.18$ &$ 2.25\porm0.14$ \\
            & 2013.119 &$  53.75\porm 0.1$ &$  94.01\porm 0.2$ &$108.24\porm0.18$ &$ 3.11\porm0.14$ \\
            & 2013.146 &$  53.77\porm 0.1$ &$  94.25\porm 0.2$ &$108.46\porm0.18$ &$ 3.22\porm0.14$ \\
\\
J1748--291  & 1996.221 &$   1.04\porm 0.2$ &$  -2.09\porm 0.6$ &$-1.24\porm0.52$ &$-1.98\porm0.36$ \\
            & 1996.252 &$   1.06\porm 0.2$ &$  -2.18\porm 0.6$ &$-1.31\porm0.52$ &$-2.04\porm0.36$ \\
            & 1997.211 &$   4.53\porm 0.2$ &$   2.76\porm 0.6$ &$ 4.72\porm0.52$ &$-2.43\porm0.36$ \\
            & 1997.241 &$   4.62\porm 0.2$ &$   2.44\porm 0.6$ &$ 4.49\porm0.52$ &$-2.68\porm0.36$ \\
            & 1998.202 &$   7.65\porm 0.2$ &$   8.34\porm 0.6$ &$11.10\porm0.52$ &$-2.19\porm0.36$ \\
            & 1998.230 &$   7.65\porm 0.2$ &$   8.10\porm 0.6$ &$10.90\porm0.52$ &$-2.31\porm0.36$ \\
            & 1999.791 &$  12.87\porm 0.2$ &$  17.79\porm 0.6$ &$21.89\porm0.52$ &$-1.72\porm0.36$ \\
            & 1999.799 &$  12.64\porm 0.2$ &$  17.18\porm 0.6$ &$21.25\porm0.52$ &$-1.84\porm0.36$ \\
            & 1999.805 &$  12.94\porm 0.2$ &$  17.84\porm 0.6$ &$21.97\porm0.52$ &$-1.76\porm0.36$ \\
            & 2000.232 &$  13.98\porm 0.2$ &$  20.05\porm 0.6$ &$24.40\porm0.52$ &$-1.48\porm0.36$ \\
            & 2000.238 &$  14.13\porm 0.2$ &$  20.54\porm 0.6$ &$24.90\porm0.52$ &$-1.36\porm0.36$ \\
            & 2000.246 &$  13.95\porm 0.2$ &$  19.90\porm 0.6$ &$24.25\porm0.52$ &$-1.54\porm0.36$ \\
            & 2003.264 &$  23.42\porm 0.2$ &$  37.89\porm 0.6$ &$44.54\porm0.52$ &$-0.25\porm0.36$ \\
            & 2003.318 &$  23.96\porm 0.2$ &$  37.17\porm 0.6$ &$44.20\porm0.52$ &$-1.09\porm0.36$ \\
            & 2007.253 &$  36.16\porm 0.2$ &$  59.41\porm 0.6$ &$69.55\porm0.52$ &$+0.09\porm0.36$ \\
            & 2007.264 &$  36.40\porm 0.2$ &$  59.60\porm 0.6$ &$69.84\porm0.52$ &$-0.02\porm0.36$ \\
            & 2007.280 &$  36.42\porm 0.2$ &$  59.58\porm 0.6$ &$69.83\porm0.52$ &$-0.05\porm0.36$ \\
            & 2013.119 &$  54.71\porm 0.2$ &$  91.93\porm 0.6$ &$106.96\porm0.52$&$+1.20\porm0.36$ \\
            & 2013.146 &$  54.82\porm 0.2$ &$  92.04\porm 0.6$ &$107.12\porm0.52$&$+1.17\porm0.36$ \\
\enddata
\tablecomments{Position offsets are relative to \sgra, after removing the
$\approx$0.7 degree differences of the background sources.  
The coordinate offsets are relative to the following J2000 positions: 
\sgrab (17 45 40.0409, --29 00 28.118),
J1745--283 (17 45 52.4968, --28 20 26.294), and 
J1748--291 (17 48 45.6860, --29 07 39.404).  
The conversion to Galactic coordinates is discussed in the Appendix of \citet{Reid:04}, and their
uncertainties have been updated to include the synthesized beam position angle in the propagation of errors.  
The positions for epochs before 1998 have been corrected for the second-order effects of processing the
phase-reference data from \sgrab with J2000 coordinates of (17 45 40.0500, --29 00 28.120).
  \label{table:positions}
}
\end{deluxetable}

\begin{deluxetable}{cccccll}
\tablewidth{0pt}
\tabletypesize{\small}
\tablecaption{Apparent Relative Motions}
\tablehead{
\colhead{Source -- Reference}         & 
\colhead{Easterly Motion}    & \colhead{Northerly Motion} &
\colhead{$\ell^{II}$ Motion}    & \colhead{$b^{II}$ Motion} &
 \\
\colhead{}               & 
\colhead{(mas y$^{-1}$)}       & \colhead{(mas y$^{-1}$)}  &
\colhead{(mas y$^{-1}$)}       & \colhead{(mas y$^{-1}$)}  &
 }
\startdata
\\
\sgra~--~J1745--2820 .......&$-3.147\pm0.008$  &$-5.578\pm0.011$ &$-6.402\pm0.010$  &$-0.219\pm0.008$ \\
\sgra~--~J1748--2907 .......&$-3.166\pm0.008$  &$-5.606\pm0.019$ &$-6.434\pm0.016$  &$-0.218\pm0.013$ \\
\\
\sgra~--~Combined    .......&$-3.156\pm0.006$  &$-5.585\pm0.010$ &$-6.411\pm0.008$  &$-0.219\pm0.007$ \\
\enddata
\tablecomments{Motions values are from weighted least-squares fits to the 
data in Table 1, with uncertainties scaled to give a reduced chi-squared of unity.  
Equatorial motions are in the J2000 system and Galactic motions are based on
Galactic coordinates transformed to J2000 as described in the appendix of 
\citet{Reid:04}.  
``Combined'' motions are variance weighted averages of the individual results.
  \label{table:motions}
}
\end{deluxetable}

\begin{deluxetable}{lrrll}
\tablewidth{0pt}
\tabletypesize{\small}
\tablecaption{Estimating \sgra's Intrinsic Motion}
\tablehead{
\colhead{Description}    & 
\colhead{\muell}    & \colhead{\mubee} &
 \\
\colhead{}               & \colhead{(\masyr)}         & \colhead{(\masyr)}  &
 }
\startdata
\sgra's {\it apparent} motion$^a$               &$-6.411\pm0.008$ 	&$-0.219\pm0.007$ \\
Reflex of Sun's Galactic orbit$^b$        &$-6.396\pm0.057$	&$-0.197\pm0.018$ \\
\\
Difference: \sgra's {\it intrinsic} motion      &$-0.015\pm0.058$	&$-0.022\pm0.019$ \\
\hline
\\
                                          &(\kms)~~~               &(\kms)~~~ \\
Difference: assuming $\rnot=8.15$ kpc     &$-0.58\phantom{0}\pm2.23\phantom{0}$       
                                          &$-0.85\phantom{0}\pm0.75\phantom{0}$  \\
\enddata

\tablenotetext{a}{Proper motion Galactic longitude (\muell) and latitude (\mubee) 
from Table \ref{table:motions}.}

\tablenotetext{b}{Adopting the Galactic orbital values of the Sun from
parallaxes and proper motions of masers associated with massive young
stars by \citet{Reid:19}.
Longitudinal motion: $(\tnot+\vsun)/\rnot = 30.32\pm0.27$ \kmskpc.
Latitudinal motion: $\wsun = 7.6\pm 0.7$ \kms.
  \label{table:intrinsicmotion}
}
\end{deluxetable}

\begin{deluxetable}{lcccc}
\tablecolumns{5} \tablewidth{0pc}
\tablecaption{Density Limits for SMBH Candidates}
\tablehead {
\colhead{Object} & \colhead{Mass}    & \colhead{Radius} & \colhead{$M/R$}   & \colhead{Density}   \\
\colhead{}       & \colhead{(\Msun)} & \colhead{(AU)}   & \colhead{(kg/m)}  & \colhead{(\Msunpercpc)} \\
           }

\startdata
Globular Cluster &$\p1\times10^6$ &$\p2\times10^5$&$\p7\times10^{19}$ &$\p3\times10^5\q$  \\
NGC~4258         &$\p4\times10^7$ &$<2\times10^4$ &$>3\times10^{22}$  &$>1\times10^{10}$  \\
Stellar orbits   &$\p4\times10^6$ &$<100$         &$>5\times10^{23}$  &$>8\times10^{15}$   \\
\sgrab  motion   &$>1\times10^6$  &$0.18$         &$>7\times10^{25}$  &$>4\times10^{23}$   \\
\\
SMBH ($3R_{Sch}$)&$\p4\times10^6$ &$0.24$         &$\p2\times10^{26}$ &$\p6\times10^{23}$  \\
\enddata

\tablecomments{The columns under the headings $M/R$ and Density give the critical 
``linear'' and volume densities, appropriate for $R_{crit} = 3 R_{Sch}$ 
as discussed in Section \ref{sect:SMBH}. 
\label{table:densities}
               }

\end{deluxetable}

\end{document}